\documentstyle[cite,epsf,12pt]{article}
\newcommand{\beq}{\begin{equation}}
\newcommand{\eeq}{\end{equation}}
\newcommand{\beqarray}{\begin{eqnarray}}
\newcommand{\eeqarray}{\end{eqnarray}}

\begin{document}


\hyphenation{Ginz-burg Le-van-yuk  com-pound multilayered 
ap-pli-ca-tion con-sid-er com-pound su-per-con-duc-ting}

\newcommand{\ie}{{i.e.}}
\newcommand{\eg}{{e.g.}}
\newcommand{\etal}{{\it et al.}}
\newcommand{\eq}[1]{~(\ref{#1})}
\newcommand{\eqdos}[2]{~(\ref{#1},\ref{#2})}
\renewcommand{\epsilon}{\varepsilon}
\newcommand{\lsim}{\stackrel{<}{_\sim}}
\newcommand{\gsim}{\stackrel{>}{_\sim}}

\newcommand{\Cha}{Chakravarty}

\newcommand{\abindex}{{\|}}
\newcommand{\cindex}{{\bot}}

\newcommand{\Tc}{\mbox{$T_c$}}
\newcommand{\TcN}{\mbox{$T_c^{N}$}}
\newcommand{\TcofN}{\mbox{$T_c(N)$}}
\newcommand{\TcNuno}{\mbox{$T_c^{N=1}$}}
\newcommand{\TcNdos}{\mbox{$T_c^{N=2}$}}
\newcommand{\TcNunoILT}{\mbox{$T_{c\;{\rm ILT}}^{N=1}$}}
\newcommand{\TcNdosILT}{\mbox{$T_{c\;{\rm ILT}}^{N=2}$}}
\newcommand{\Tcab}{\mbox{$T_{c\abindex}$}}

\newcommand{\cjump}{\mbox{$c_{\rm jump}$}}
\newcommand{\cjumpN}{\mbox{$c_{\rm jump}^N$}}

\newcommand{\lc}{\mbox{$\lambda_c$}}
\newcommand{\Econd}{\mbox{$E_0$}}
\newcommand{\EcondILT}{\mbox{$E_0^{\rm ILT}$}}

\newcommand{\DF}{\mbox{$\Delta f$}}
\newcommand{\DFab}{\mbox{$\Delta f_\abindex$}}
\newcommand{\DFcint}{\mbox{$\Delta f_{\cindex{\rm int}}$}}
\newcommand{\DFcext}{\mbox{$\Delta f_{\cindex{\rm ext}}$}}

\newcommand{\DFFsinF}{\mbox{$\Delta F$}}
\newcommand{\DFF}{\mbox{$\Delta F[\Psi]$}}
\newcommand{\DFFo}{\mbox{$\Delta F[\Psi_{jn}(0)]$}}
\newcommand{\Fo}{\mbox{$\Psi_{jn}(0)$}}

\newcommand{\xico}{\mbox{$\xi_c(0)$}}

\newcommand{\gammaint}{\mbox{$\gamma_{\rm int}$}}
\newcommand{\deltaint}{\mbox{$\delta_{\rm int}$}}
\newcommand{\gammaext}{\mbox{$\gamma_{\rm ext}$}}
\newcommand{\deltaext}{\mbox{$\delta_{\rm ext}$}}
\newcommand{\gammaintext}{\mbox{$\gamma_{\rm int,ext}$}}
\newcommand{\deltaintext}{\mbox{$\delta_{\rm int,ext}$}}

\newcommand{\fcint}{\deltaint}
\newcommand{\fcext}{\deltaext}
\newcommand{\fcintext}{\deltaintext}

\newcommand{\epsilonabT}{\mbox{$\epsilon_\abindex(T)$}}
\newcommand{\epsilonabTc}{\mbox{$\epsilon_\abindex(\Tc)$}}
\newcommand{\epsilonabTcN}{\mbox{$\epsilon_\abindex(\TcN)$}}
\newcommand{\eptilde}{\mbox{$\tilde\epsilon$}}
\newcommand{\eptildeT}{\mbox{$\tilde\epsilon(T)$}}
\newcommand{\eptildeTc}{\mbox{$\tilde\epsilon(\Tc)$}}
\newcommand{\eptildeTcN}{\mbox{$\tilde\epsilon(\TcN)$}}

\newcommand{\CuOdos}{CuO$_2$}
\newcommand{\Tluno}{Tl$_1$}
\newcommand{\Tldos}{Tl$_2$}
\newcommand{\Hg}{Hg}
\newcommand{\La}{La}
\newcommand{\TlunoNuno}{Tl$_1^{N=1}$}
\newcommand{\TldosNuno}{Tl$_2^{N=1}$}
\newcommand{\HgNuno}{Hg$^{N=1}$}
\newcommand{\LaNuno}{La$^{N=1}$}
\newcommand{\BSCOf}{Bi$_2$Sr$_2$CaCu$_2$O$_{8+x}$}
\newcommand{\BSCO}{Bi-2212}
\newcommand{\YBCOf}{YBa$_2$Cu$_3$O$_{7-\delta}$}
\newcommand{\YBCO}{Y-123}

\newcommand{\equis}{\mbox{\textsf{A}}}


\newcommand{\titulo}{
On the energy saved by interlayer interactions in the superconducting state of cuprates
}

\newcommand{\autor}{Manuel V.~Ramallo\footnote{fmrama@usc.es}}

\newcommand{\direccion}{
Department of Physics, University of Illinois,\\ 1110 W.~Green Street, Urbana, Illinois 61801-3080, USA,\vspace{3pt}\\ 
and
Laboratorio de Baixas Temperaturas e
Superconductividade\\  (Unidad Asociada al ICMM-CSIC,
Spain),\\  Departamento de F\'{\i}sica da Materia
Condensada,\\  Universidade de Santiago de Compostela,
E-15782, Spain\footnote{Present and permanent address, to where correspondence should be addressed}}

\hrule  
\begin{center}
\mbox{}\vspace{-0.5cm}\\
{\nonfrenchspacing\it Revised~15~January~2004~(as~published~in~Europhys.~Lett.~{\bf 65},~249-255~(2004)\,)}\vspace{0.3cm}\\ 
{\it First version 21 August 2003: http://arxiv.org/abs/cond-mat/0308423v1}\\ 
 \end{center}
  \hrule \mbox{}\\
\mbox{}\vspace{-1cm}\\ 
\begin{center}
  \Large\bf
\titulo\\  \end{center}\mbox{}\vspace{-1cm}\\

\begin{center}\normalsize\autor\end{center} 

\begin{center}\normalsize\it\direccion\end{center}


\mbox{}\vskip0.5cm{\bf Abstract. }
A Ginzburg-Landau--like functional is proposed reproducing the main low-energy features of various possible high-\Tc\ superconducting mechanisms involving energy savings due to interlayer interactions. The functional may be used to relate these  savings to experimental quantities. Two examples are given, involving  the mean-field specific heat jump at \Tc\ and the superconducting fluctuations above \Tc. Comparison with existing data suggests, \eg, that the increase of \Tc\ due to the so-called interlayer tunneling (ILT) mechanism of interlayer kinetic-energy savings is negligible in optimally-doped \BSCOf.

\newpage
\setlength{\baselineskip}{18pt}


\section{Introduction}
One of the striking systematics of the superconducting critical temperature \Tc\ of the cuprate superconductors (HTSC) is its correlation, within each family of chemically similar compounds,  with the number $N$ of \CuOdos\ planes per unit cell.\cite{exper-TcN,ILT,MIR} Well-known examples of these families are the Ca-spaced series, with formula {\equis}Ca$_{N-1}$(\CuOdos)$_N$, such as the so-called \La-, \Hg-, \Tluno-, and \Tldos-series (where \equis\ is, respectively, La$_{2-x}$Sr$_x$O$_2$,  HgBa$_2$O$_2$, TlBa$_2$O$_{3-\delta}$, and Tl$_2$Ba$_2$O$_{4-\delta}$). In all these Ca-spaced series, the following rule is observed to be well obeyed,\cite{exper-TcN,ILT,MIR} at least for low $N$ values, $N\leq3$\cite{N4}:
\beq
\frac{\TcN-\TcNuno}{\TcNuno}\propto\left(1-\frac{1}{N}\right)\,,
\label{Tc(N)}
\eeq
where \TcNuno\ and \TcN\ are the critical temperatures of, respectively, the single-layered and the $N$-layered compounds (at zero magnetic field and for bulk samples with optimal hole doping concentration).  The origin of relation\eq{Tc(N)} could in principle be sought either in some $N$-dependence of the parameters involved in in-plane interactions leading to superconductivity, or in the existence of  $c$-direction (\ie, inter-layer) interactions contributing, at least in part, to the condensation energy. The latter way of thinking is consistent with the proposals made by various authors of different forms of interlayer interactions in HTSC saving energy in the superconducting state.\cite{otros,ILT,MIR} For instance, eq.\eq{Tc(N)} was obtained, for low $N$,  by Leggett\cite{MIR,N4} by considering the screening of Coulomb interlayer interactions among carriers. Energy savings in the superconducting state occur in this approach mainly in the potential energy of the electrons.\cite{MIR} In what concerns eq.\eq{Tc(N)} and for low $N$, Leggett's formalism may be expected to apply {\it in essence} for a broad variety of superconducting mechanisms responsible for the $c$-direction attractive screening.\cite{MIR} A notable exception  is the so-called interlayer tunneling (ILT) mechanism proposed by Anderson,  \Cha, and coworkers\cite{ILT}.  In the ILT model, 
savings occur in the  kinetic energy due to a deconfinement process of the Cooper pairs. This is originated by strong electronic correlations that block the coherent interlayer tunneling for single particles, but not for pairs. Strikingly, the ILT proposal again leads to eq.\eq{Tc(N)} in spite of the very different origin of the interlayer energy savings.\cite{KT}

An experimental test was proposed by Anderson\cite{AndersonLambda} and Leggett\cite{LeggettLambda} to estimate in single-layered (\ie, $N=1$) HTSC the importance of the ILT mechanism. The test involves the condensation energy \Econd, obtained from specific heat data, and the $c$-direction magnetic penetration depth at $T=0$~K. Anderson\cite{AndersonLambda} argued that data existing for the $N=1$ compounds of the \La-series\cite{testLaSi} and \Hg-series\cite{testHgSi} agreed with the ILT prediction. In contrast, subsequent measurements of \lc\ in the $N=1$ compounds of the  \La-series,\cite{testLaTldosNo} \Hg-series,\cite{testHgNo} and \Tldos-series,\cite{testLaTldosNo,testTldosNo} concluded that  the ILT mechanism gives a negligible contribution to \Econd. \Cha\ \etal\cite{Cha} then pointed that, in addition to these discrepancies between different measurements, the tests could be affected by ambiguities in the obtainment of \Econd; this aspect has been recently answered in part in\cite{respuestaCha}. Note that, because of the $N=1$ limitation, the above tests have probed only the energy savings due to the interaction between layers in different unit cells. However, this extra-cell interaction may be expected to be the less significative one for enhancing \Tc; therefore, the failure of a given superconducting interlayer mechanism to account for the condensation energy of the $N=1$ cuprates does not rule it out completely as a substantial source for the enhancement of \Tc\ when $N>1$ (see also below).

In the present work, we introduce a Ginzburg-Landau (GL)--type functional that reproduces in essence the main low-energy features of the above proposals of interlayer kinetic and/or potential energy savings in HTSC. This functional is based on a simple energy-balance argument expected to be a good approximation near the transition and for low Cooper-pair densities, and may be useful for finding experimentally testable relationships involving the interlayer superconducting energy savings in HTSC with $N\geq1$. Two examples of such tests will be given in this letter: The first is a relationship between $N$ and the mean-field specific heat jump at \Tc. If it is fulfilled in a family of chemically similar HTSC, it will indicate that the superconductors in the series differ only in their interlayer interactions (\ie, that eq.\eq{Tc(N)} is actually due to a superconducting interlayer mechanism). The second test will be provided by the Gaussian superconducting fluctuations above \Tc: Our present functional leads, for zero or weak magnetic fields,  to fluctuations identical to  the well-known ones of multilayered superconductors with no interlayer energy savings; however, the involved parameters acquire now additional meanings. In particular, the interlayer Josephson tunnelings are related to the maximum increase of \Tc\ that could be attributed to the ILT mechanism.  Finally, we end this letter with a brief  discussion of available experimental data.

\section{A preliminary energy-balance argument leading to eq.\eq{Tc(N)}} 
In a layered superconductor with $N$ superconducting layers in each $c$-direction unit cell of length $s$, we may write the free energy \DF\ saved in the superconducting state, per Cooper pair and unit volume, as the following sum:
\beq
\DF=s^{-1}\left[N\DFab+(N-1)\DFcint+\DFcext\right]\,.
\label{bal-1}
\eeq
Here \DFab, \DFcint\ and \DFcext\ are the energy savings, per Cooper pair and unit area, due to, respectively, the in-plane interactions in each superconducting layer, the interactions between each two adjacent layers in the same cell (intra-cell interaction), and the interactions between each two adjacent cells (extra-cell interaction). We assumed in eq.\eq{bal-1} that all the superconducting layers and intra-cell separations between them are equivalent. We also assumed that the energies saved by each interlayer interaction may be considered as independent, and that the interactions between non-adjacent layers are negligible. These assumptions are expected to be good approximations in both Leggett's and ILT proposals.\cite{ILT,MIR} Let us now expand \DFab\ in powers of the reduced temperature $\epsilonabT\equiv\ln(T/\Tcab)\simeq(T-\Tcab)/\Tcab$, where \Tcab\ is the critical temperature of the system if the interlayer interactions were absent; this  expansion is obviously reminiscent of the type of reasonings used in the GL-like theories. So, for the in-plane interactions above $\epsilonabT=0$ we write $\DFab=-\alpha\epsilonabT$, where $\alpha$ is  constant and positive. For the interlayer interactions, as they will be less dependent on \epsilonabT, we may write in first approximation $\DFcint=\alpha\fcint$ and $\DFcext= \alpha\fcext$,  where \fcint\ and \fcext\ are dimensionless constants (positive, if the corresponding interactions help superconductivity). After these simple power expansions, the actual critical temperature \TcN\ can be easily calculated by just writing  the condition $\DF(\TcN)=0$, \ie,
\beq
\epsilon_{\rm eff}^N(\TcN)=0, \qquad\mbox{with}\qquad
\epsilon_{\rm eff}^N(T)\equiv\epsilonabT-\frac{\fcext}{N}-\left(1-\frac{1}{N}\right)\fcint\,.
\label{bal-2}
\eeq
which directly leads to:
\beq
\label{bal-4}\frac{\TcN-\TcNuno}{\TcNuno}  \simeq \ln\frac{\TcN}{\TcNuno} =  \left(\fcint-\fcext\right)\;\left(1-\frac{1}{N}\right)\,.
\eeq
We find, therefore, a result equivalent to eq.\eq{Tc(N)}, independently of whether the mechanism of interlayer energy savings imply the kinetic or potential energies, or both. In the above equation we have considered it useful to explicitly emphasize the fact that the logarithm of the quotient of two temperatures is approximately equal to the relative distance between them; we will omit the explicit emphasis of this point in the remainder of this paper.

Note that for $N=1,2$ the above equations also lead to:
\beq
\label{bal-5}\ln\frac{\TcNuno}{\Tcab}  = \deltaext\,,\qquad
\ln\frac{\TcNdos}{\Tcab} = \frac{\deltaext+\deltaint}{2}\,.
\eeq
These relationships indicate that if $\deltaext\ll\deltaint$  is possible for $\TcNdos$ to be quite different from $\TcNuno$ even if $\TcNuno\simeq\Tcab$. In this respect, we note that the recent observation of superconductivity in a sample of the $N=1$ La-series compound with thickness only one $c$-direction unit cell\cite{Bozovic} is not contradictory with the existence of interlayer superconducting energy savings in HTSC.  These aspects also enhance the interest of testing the existence or not of interlayer superconducting energy savings in the $N>1$ compounds.

\section{A simple Ginzburg-Landau (GL) free-energy functional for HTSC with interlayer energy savings} Our next step is to introduce a GL-type functional consistent with the main proposals for interlayer kinetic and/or potential energy savings in HTSC. This functional must be chosen so that for low energies and Cooper-pair densities it recovers the energy balances of the previous section. This is fulfilled by the following expression for the difference between the superconducting and normal-state free energies, \DFF, at zero magnetic field:
\beqarray
\DFF&=& 
\int{\rm d}^2{\bf r}\sum_{n}
\sum_{j=1}^{N} a_0
\left\{\;
\epsilonabT|\Psi_{jn}|^2+
\frac{\mbox{$b$}}{\mbox{$2a_0$}}|\Psi_{jn}|^4+
\xi_{ab}^2(0)|\nabla_{xy}\Psi_{jn}|^2\right.\nonumber\\
&&\left.-\gamma_j \left(\Psi_{jn}\Psi_{j+1,n}^*+{\rm c.c.}\right)+(2\gamma_j-\delta_j)\;\left(
\frac
{|\Psi_{jn}|^2+|\Psi_{j+1,n}|^2}
{2}
\right)\;
\right\}\,.
\label{DFGL}
\eeqarray
In the above expression, ${\bf r}=(x,y)$ and $\nabla_{xy}$ are the in-plane coordinates and gradient (we neglect the possible in-plane anisotropy); the indexes $(j,n)$ label each $j$th superconducting plane of the  $n$th $c$-direction unit cell (we use also $(N+1,n)$ for the $(1,n+1)$ plane); $\Psi_{jn}=\Psi_{jn}(x,y)$ is the superconducting order parameter of the $(j,n)$-plane; $\xi_{ab}(0)$ is the GL amplitude of the in-plane coherence length; $a_0$ and $b$ are the GL constants. The interlayer interactions in the above functional are contained in the last two summands. In them, $\gamma_j$ is the usual Josephson coupling constant between the $(j,n)$ and $(j+1,n)$ planes, and $\delta_j$ parametrizes the interlayer energy savings due to interactions also between these two planes. These $\delta_j$ parameters coincide with the ones already used in eqs.\eq{bal-2}~--\eq{bal-4}, and to our knowledge  this is the first time that they are introduced in a GL model of a layered superconductor.   As we did in the previous section, we may assume in the HTSC only two different interlayer separations (intra- and extra-cell) and so we take $\gamma_N=\gammaext$, $\delta_N=\deltaext$ and $\gamma_{j\neq N}=\gammaint$, $\delta_{j\neq N}=\deltaint$.

The parameters $\delta_j$ deserve further discussion. When they are zero, the interlayer summands in eq.\eq{DFGL} reduce to $\gamma_j|\Psi_{jn}-\Psi_{j+1,n}|^2$. As the latter expression is never negative, it does not lead to energy savings. In fact, this  is precisely the well-known  expression for the interlayer interaction in the extended--Lawrence-Doniach GL-functional for multilayered superconductors without interlayer energy savings.\cite{Klemm,Ne,cfl} In contrast, when $\delta_j>0$ the interlayer energy can be negative: For instance, if $\delta_j=2\gamma_j$ the last interlayer summand in eq.\eq{DFGL} is zero and the first one becomes proportional to $\cos\varphi_j$, where $\varphi_j$ is the difference of the phases of $\Psi_{jn}$ and $\Psi_{j+1,n}$. This is precisely the form of the interlayer kinetic-energy savings proposed on microscopical grounds by the ILT model.\cite{ILT,AndersonLambda} 
 Finally, interlayer potential energy savings as those in Leggett's proposals for the superconducting mechanism\cite{MIR} may also be crudely included in our functional by considering different $\delta_j>0$. This is coherent with the fact that these savings may be expected to arise in the second of the interlayer interaction summands, \ie, the one not involving the phase of the superconducting wave function. Let us also note that  these kinetic and potential interlayer energy savings could coexist in the HTSC; in that case, the $\delta_j$ would result from the sum of the contribution of each energy-saving source, leading then to energy savings essentially of the form $A+B\cos\varphi_j$, as in fact proposed in \cite{LeggettLambda} for that mixed scenario.

It is not difficult to calculate the \TcN\ resulting from eq.\eq{DFGL}. For that, one just has to write the condition $\DFFo(\TcN)=0$, where \Fo\ are the equilibrium wave functions, verifying $\partial\DFFsinF/\partial\Psi_{jn}=0$. Just as expected, \TcN\ simply follows eqs.\eq{bal-2}~--\eq{bal-5}. 

\section{Jump at \Tc\ of the mean-field specific heat} It is also quite direct to calculate from eq.\eq{DFGL} the mean-field specific heat jump at the transition, \cjumpN, \ie, the discontinuity at $T=\TcN$ in the specific heat when considering the {\it uniform} wave function minimizing \DFF. We get:
\beq
\cjump=
\frac{N}{s\TcN}\,\frac{a_0^2}{b}\,,
\label{cj}
\eeq
where again $s$ is the layered-structure repetition distance.
This proportionality  of \cjumpN\ and $N/s\Tc$ has been to the best of our knowledge unnoticed up to now even for layered superconductors with no interlayer energy savings. Two important additional remarks are: First, \cjumpN\ is found to be independent of the interlayer interaction parameters, both $\delta_j$ and $\gamma_j$ (except indirectly trough \TcN, see eq.\eq{bal-4}). Second, according to eq.\eq{cj} the quantity $\cjump/(N/s\Tc)$ will be the same for all the HTSC with equal in-plane interactions (and hence equal GL parameters $a_0$ and $b$). In particular, experimental verification of this constancy in a series of chemically similar HTSC would provide a compelling argument favoring that eq.\eq{Tc(N)} is in fact due {\it only} to the existence of  a superconducting interlayer mechanism, and not to some $N$-dependence of the parameters involved in in-plane interactions. (We note however that {\it failure} of fulfilment of that proportionality would not rule out completely interlayer interactions contributing at least in part to the \Tc\ enhancement, as they could still coexist with in-plane effects varying $a_0$ and $b$ in the series.)

\section{Gaussian-Ginzburg-Landau (GGL) fluctuations\\ above \Tc} Let us now consider the superconducting fluctuations above \Tc\ that result from  functional\eq{DFGL}. For that, we apply to it the same formalism previously used in \cite{Ne,cfl} to study the Gaussian regime of the superconducting fluctuations in  multilayered superconductors without interlayer energy savings, \ie, with $\delta_j=0$. The calculations go in parallel to those in~\cite{Ne,cfl}, and so we will not make them explicit in this letter. The important result is that the fluctuation spectrum of functional\eq{DFGL} with arbitrary $\delta_j$ is just the same as found in~\cite{Ne,cfl} for  $\delta_j=0$, {\it if} we interpret in these equations \Tc\  as \TcN. This applies also to the results directly derived from that spectrum, including the fluctuation-induced observables  calculated in~\cite{Ne,cfl} at $T>\Tc$ for zero magnetic fields and also for weak magnetic fields perpendicular to the layers, \ie,    the in-plane paraconductivity\cite{Ne},  the in-plane magnetoconductivity\cite{Ne}, the fluctuation susceptibility\cite{Ne},  and the fluctuation specific heat\cite{cfl}. Therefore, for all these observables we conclude that {\it the final expressions obtained in refs.\cite{Ne,cfl}}~~(\ie, eqs.(3.4)--(3.8) and (4.5)--(4.7) in ref.\cite{Ne}, and (9)--(12) in ref.\cite{cfl}) {\it are applicable for all values of $\delta_j$}.

The above result indicates that any analysis of experimental data based on the above equations of refs.\cite{Ne,cfl} will be also applicable in the $\delta_j\neq0$ scenario. This is an important information because those analyses are able to determine rather unambiguously the values of the Josephson couplings between layers, \gammaint\ and \gammaext\ (see \cite{Ne,cfl,tau,BSCO,review} and also below). 
Precisely these Josephson couplings give a {\it maximum} limit for the interlayer kinetic energy savings in eq.\eq{DFGL}. In particular,   the maximum relative increase of \Tc\ that may be due to the ILT mechanism is given by eq.\eq{bal-5} with $\deltaint=2\gammaint$ and $\deltaext=2\gammaext$. Note that in \cite{Ne,cfl,tau,BSCO,review} instead of the variables \gammaint\ and \gammaext\ it is often used the equivalent set composed by \xico\ and \gammaint/\gammaext, where \xico\ is the $c$-direction GL coherence length amplitude; in terms of these alternative variables, eq.\eq{bal-5} leads to a maximum increase of \Tc\ due to the ILT mechanism of: 
\beqarray
\label{TcILTNuno} \ln\frac{\TcNunoILT}{\Tcab}&
\leq&2\left(\frac{\xico}{s}\right)^2,\\
\label{TcILTNdos}\ln\frac{\TcNdosILT}{\Tcab} & \leq&2\left(\frac{\xico}{s}\right)^2\left(1+\frac{\gammaext}{\gammaint}\right)\left(1+\frac{\gammaint}{\gammaext}\right)\,.
\eeqarray

\section{A brief comparison with existing experimental information} 
Testing  eq.\eq{cj} in a given family of HTSC requires reliable knowledge of \cjump\ on it. Unfortunately, the latter proves difficult. An illustrative example of these difficulties happens in the Tl$_2$-series.  The $N=2$ and 3 compounds of this series were measured by the same group, in similar samples, and using the same criterion for obtaining \cjump, in \cite{Junod}. The corresponding   $\cjump/(N/s\Tc)$ are equal to $2.6\times10^{-3}$~J/m$^2$ for $N=2$, and $2.8\times10^{-3}$~J/m$^2$ for $N=3$. This suggests fulfillment of eq.\eq{cj}, \ie, that the $\Tc(N)$ variation in this series is due {\it only} to interlayer interactions. However, measurements by other group\cite{Loram} in the $N=1$ compound of the same series revealed a quite small and symmetric specific heat peak around \Tc. This may be interpreted as a negligible \cjump\ (and then not fulfillment of eq.\eq{cj}), or could have its origin in, \eg, \Tc-inhomogeneities broadening and symmetrizing the specific heat peak around \Tc\ \cite{Houston}, or in pseudogap-induced effects\cite{Loram}. Therefore, an ultimate testing of eq.\eq{cj} would imply a more extensive analysis of the experimental specific heat. Naturally, it should also include other Ca-spaced series. Such a detailed study is beyond the scope of the present letter.

Much less ambiguity exists at present in understanding the  superconducting fluctuations above \Tc\ in HTSC  in terms of the GGL model of multilayered superconductors. As commented above, for our present purposes the main interest lies in the $N>1$ compounds. Measurements in high-quality single-crystal samples are available for at least two bilayered and optimally doped HTSC,  \BSCOf\ (\BSCO) and \YBCOf\ (\YBCO). Those data were extensively analyzed in terms of the GGL model of multilayered superconductors with null interlayer energy savings.\cite{Ne,cfl,tau,BSCO,review}  As mentioned above, these kinds of analyses remain fully applicable in the case of nonzero interlayer energy savings, with the bonus that the obtained Josephson coupling parameters give the maximum increase of \Tc\ that could be attributed to the ILT mechanism. For instance, in \cite{Ne,cfl,review} it was determined for \YBCO\ that  $\xico\simeq1.1\pm0.1$~\AA, and $1/30\lsim\gammaint/\gammaext\lsim30$. By using eq.\eq{TcILTNdos}, these values suggest an {\it upper limit} of around 70\% for the increase of \Tc\ due to kinetic-energy savings. In the case of \BSCO, the results are more conclusive: The analysis of the superconducting fluctuations in this compound leads to  $\xico\lsim0.5$~\AA, and to $\gammaint/\gammaext$ values whose boundaries  compatible with experiments depend on \xico\ roughly as $\xi_c^2(0)\lsim\gammaint/\gammaext\lsim\xi_c^{-2}(0)$ (where \xico\ is in \AA).\cite{BSCO,review} When these values are used in eq.\eq{TcILTNdos}, they lead to around 1\% maximum increase of \Tc\ due to the ILT mechanism in \BSCO.

\section{Conclusions} We introduced a simple Ginzburg-Landau--type functional that reproduces the main low-energy features of the existing proposals of interlayer kinetic and/or potential energy savings in HTSC. This functional may be used to find relationships between these savings and experimental observables. Two examples of such relationships were proposed: The first involves the mean-field specific heat jump at \Tc, \cjump, and the normalized mean interlayer distance $s\Tc/N$; if both are inversely proportional to each other in a series of HTSC, this would indicate that they share identical in-plane interactions and their \Tc's are different due only to interlayer interactions. The second relationship involves the superconducting fluctuations above \Tc\ at zero or weak magnetic fields, from which it may be obtained a maximum limit for the relative increase of \Tc\ due to  interlayer {\it kinetic} energy savings (as those in the ILT model). When compared with available experiments, this second relationship indicates  that the increase of \Tc\ due to interlayer kinetic energy savings is negligible in optimally-doped \BSCO, and between zero and $\sim70\%$ in optimally-doped \YBCO. Although based on a simplified model, these conclusions can be expected to be, at least, qualitatively correct. 
They provide, to the best of our knowledge, the first test of the significance of the interlayer kinetic energy savings in the \Tc\ of any $N>1$ HTSC. 

\mbox{}

\mbox{}

\mbox{}\\ {\Large \bf Acknowledgements}\\ \mbox{}\\
\nonfrenchspacing
The main part of this research was performed in the Department of Physics of the University of Illinois at Urbana-Champaign thanks to support from the Fulbright Foundation and to the hospitality of Prof. A.J. Leggett, who first called my attention to the problem of the interlayer condensation energy in cuprates and with whom I enjoyed various valuable talks on the subject. I am also grateful to Prof. F. Vidal for many enlightening discussions about the superconducting fluctuations and the jump at \Tc\ of the specific heat. Financial support is also acknowledged to the CICYT, Spain, under grant No. MAT2001-3053.

\newpage

 \end{document}